\documentclass[annual]{acmsiggraph}

\TOGonlineid{}

\TOGvolume{}
\TOGnumber{}

\TOGarticleDOI{}

\TOGprojectURL{}
\TOGvideoURL{}
\TOGdataURL{}
\TOGcodeURL{}

\title{A Reciprocal Formulation of Nonexponential Radiative Transfer.   3: Binary Mixtures}

\author{ Eugene d'Eon \\
8i }

\pdfauthor{ Eugene d'Eon }


\usepackage{smrdefaults}

\usepackage{color}

\usepackage{multirow}

\usepackage{rotating}
\newcommand{\mfp}{\langle s \rangle}

\newcommand{\s}{\langle s_c \rangle}  
\renewcommand{\ss}{\langle s_c^2 \rangle}  
\newcommand{\sss}{\langle s_c^3 \rangle}  
\newcommand{\ssss}{\langle s_c^4 \rangle}  

\newcommand{\su}{\langle s_u \rangle}

\newcommand{\sigmean}{\langle \Sigma \rangle}  
\newcommand{\sigtilde}{\tilde{\Sigma}}

\newcommand{\e}{\text{e}}

\def\rm{R^{-}}

\usepackage{paralist}
\usepackage{subfigure}
\usepackage{rotating}
\usepackage{amsgen}
 
\usepackage{alltt}
\usepackage{ulem}
\usepackage{xcolor}

\begin{document}

\maketitle

\begin{abstract}
  We derive the form of reciprocal generalized radiative transfer (RGRT) that includes the Levermore-Pomraning attenuation law for paths leaving a deterministic origin.  The resulting model describes linear transport within multi-dimensional stochastic binary mixtures with Markovian mixing statistics and nonstochastic albedo and phase function.  The derivation includes a new attenuation law and related free-path distribution between collisions, which was previously estimated using a Monte Carlo approach. 
  We also show that previous stationary descriptions of binary mixtures with binomial mixing statistics reduce to exponential attenuations and, thus, have an exact homogenization within classical radiative transfer.
\end{abstract}

\keywordlist

\section{Intro}

  \normalem

  In the previous two papers in this series~\cite{deon2018reciprocali,deon2018reciprocalii} we presented motivation and details of a formulation of reciprocal generalized radiative transfer (RGRT) that accounts for correlation between scattering events in piecewise homogeneous media.  When scatterers in random media are spatially correlated, the chord-length distributions between them are nonexponential.  Classical transport theory is extended to account for the nonexponential free-path lengths between scatterers by using a semi-Markov nonexponential random flight model with a two-point memory~\cite{grosjean51}.  The nonclassical Boltzmann equation that describes such transport has recently been derived~\cite{larsen11}.  

  Audic and Frisch~\shortcite{audic1993monte} proposed the first reciprocal form of two-point nonclassical transport in the context of stochastic binary mixtures.  They noted the necessary distinction between free-path length statistics for a path beginning at the boundary and the free-path lengths between collisions.  They proposed accumulating histograms of the intercollision free-path lengths within a given class of binary mixtures by a preprocess of tracking Monte Carlo histories in explicit realizations of that class.  We show that RGRT naturally predicts the intercollision free path distribution directly from the Pomraning-Levermore attenuation law, and we find this prediction to agree with Monte Carlo simulation in a rod with forward scattering.  

  The resulting form of RGRT should improve the utility of Audic and Frisch's accelerated Monte Carlo approach to binary mixtures, which should be more efficient than the chord-length sampling (CLS) method~\cite{zimmerman1991algorithms,donovan2003application,larmier2018monte} and should be useful for predicting transport in molecular clouds~\cite{boisse1990radiative}, shielding materials~\cite{becker2014measurement}, and other applications of clumpy stochastic media~\cite{pomraning98,sanchez91}.

  \section{Markovian Binary Mixtures}

    In this section we derive the statistical functions required by RGRT to describe transport in Markovian binary mixtures with nonstochastic albedo and nonstochastic phase function.  This includes attenuation laws $X_c$, $X_u$, and related free-path distributions $p_c$, and $p_u$, where ``c'' refers to free paths with an origin that is \emph{correlated} spatially to scatterers in the medium.  Deterministic origins, such as boundary interfaces and imbedded objects use the label ``u'' to denote their \emph{uncorrelated} relationship to the scatterers.  These quantities and their relationships are summarized in Table~\ref{tab:notation}.

    The exact, uncorrelated-origin attenuation law for binary mixtures with Markovian mixing statistics is~\cite{avaste1974solar,levermore86,vanderhaegen1986radiative}
  \begin{equation}
    X_u(s) = \frac{r_+ - \sigtilde}{r_+ - r_-} \e^{-r_+ s} + \frac{\sigtilde - r_-}{r_+ - r_-} \e^{-r_- s}
  \end{equation}
  where the decay constants are
  \begin{equation}
    2 r_\pm = \sigmean + \sigtilde \pm \sqrt{\left( \sigmean - \sigtilde \right)^2+4 \beta }
  \end{equation}
  with parameters
  \begin{align}
    &\sigmean = p_A \Sigma_A + p_B \Sigma_B \\
    &\sigtilde = p_B \Sigma_A + p_A \Sigma_B + \lambda_A^{-1} + \lambda_B^{-1} \\
    &\beta = (\Sigma_A - \Sigma_B)^2 p_A p_B \\
    &p_i = \frac{\lambda_i}{\lambda_A + \lambda_B}.
  \end{align}
  The medium is parameterized by the macroscopic cross-sections for the two phases, $\Sigma_i \ge 0$, and the mean chord lengths within each phase, $\lambda_i > 0$.  Using the volume fractions $p_i$ for each phase yields the atomic mix total macroscopic cross-section $\sigmean$.

\begin{figure}
      \centering
      \includegraphics[width=.9 \linewidth]{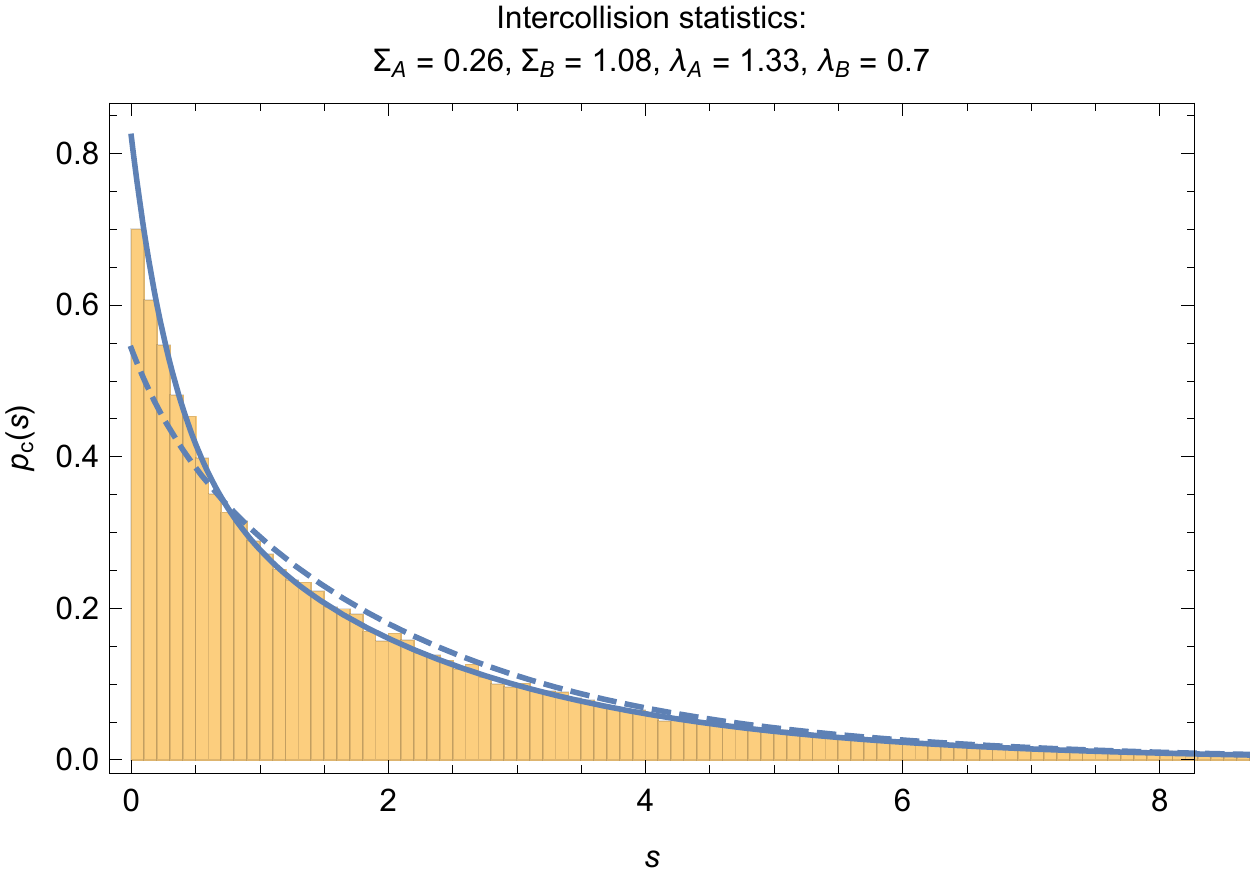}
      \caption{Distribution of intercollision free-path lengths $p_c(s)$ for a binary stochastic medium with Markovian mixing statistics.  Monte Carlo (filled) vs Eq.(\ref{eq:pc}) (continuous).  The known exact free-path distribution of distance-to-first-collision from entry at the boundary ($p_u(s)$) is shown (dashed) for reference.  The Monte Carlo simulation averages 60,000 histories in 60,000 unique realizations, each with one first collision and one second collision (assuming forward scattering in a rod).}
      \label{fig:MC} 
  \end{figure}

  To produce reciprocal transport under this attenuation law in bounded media with scattering and nonstochastic single-scattering albedo $c$ requires~\cite{deon2018reciprocali} that the distribution of free-path lengths between collision events is the normalization of
  \begin{equation}
    \frac{\partial^2}{\partial s^2} X_u(s) = \frac{r_+ - \sigtilde}{r_+ - r_-} r_+^2 \e^{-r_+ s} + \frac{\sigtilde - r_-}{r_+ - r_-} r_-^2 \e^{-r_- s}.
  \end{equation}
  We find that the inverse of the normalization constant is
  \begin{equation}
    \int_0^\infty \frac{\partial^2}{\partial s^2} X_u(s) ds = r_+ + r_- - \sigtilde = \sigmean
  \end{equation}
  yielding the correlated-origin free-path distribution
  \begin{equation}\label{eq:pc}
    p_c(s) = \frac{1}{\sigmean} \left( \frac{r_+ - \sigtilde}{r_+ - r_-} r_+^2 \e^{-r_+ s} + \frac{\sigtilde - r_-}{r_+ - r_-} r_-^2 \e^{-r_- s} \right)
  \end{equation}
  with the mean correlated mean free path
  \begin{equation}\label{eq:sc}
    \s = \int_0^\infty p_c(s) ds = \frac{1}{\sigmean}.
  \end{equation}
  This probability distribution function $p_c(s)$ is the ensemble-averaged distribution of distances between collisions over arbitrarily many phase transitions.  We are not aware of this result being published previously and note that a simple Monte Carlo experiment similar to~\cite{adams89} shows excellent agreement with Eq.(\ref{eq:pc}) (Figure~\ref{fig:MC}).  In this experiment, we sampled explicit realizations of 1D rod mixtures, entering at the boundary, scattering forward with no absorption and accumulating the depths of collisions 1, 2, and 3.  We found excellent agreement between $p_u(s)$ and the location of the first collision.  Likewise, for the statistics of distances between collisions 1 and 2, and also between 2 and 3, we found excellent agreement with $p_c(s)$.

  Eq.(\ref{eq:sc}) predicts that the mean free path between collisions in a Markovian binary mixture is invariant to the mixing parameters and is simply the classical homogeneous mean free path produced by the atomic mix assumption.  

  \begin{table*}[t]
    \centering     
    \scalebox{0.7}{
    \begin{tabular}{| c | l | c |} 
    
      \hline    
      \vline width0pt height2.3ex
      \textbf{Symbol} & \textbf{Description} & \textbf{Relations} \\
      \hline 
      \multicolumn{3}{|c|}{ \small \emph{medium-\textbf{c}orrelated (stochastic) free path origins }} \\
      \hline
      $s$ & {\small distance since last medium collision or correlated birth} & \\
      \hline
      $\Sigma_{tc}(s)$ & {\small correlated macroscopic cross section} & $\Sigma_{tc}(s) = \frac{p_c(s)}{X_c(s)}$ \\
      \hline
      $p_c(s)$ & {\small correlated free-path distribution} & $p_c(s) = \Sigma_t(s) \e^{-\int_0^s \Sigma_t(s') ds'} = - \frac{\partial}{ \partial s} X_c(s) = \s \frac{\partial^2}{ \partial s^2} X_u(s)$ \\
      \hline
      $X_c(s)$ & {\small correlated-origin transmittance} & $X_c(s) = 1 - \int_0^s p_c(s') ds'$ \\
      \hline
      $\mfp_c$ & {\small mean correlated free-path} & $\mfp_c = \int_0^\infty p_c(s) \, s \, ds$ \\
      \hline
      $\ss$ & {\small mean squared correlated free-path} & $\ss = \int_0^\infty p_c(s) \, s^2 \, ds$ \\
      \hline
      \multicolumn{3}{|c|}{ \small \emph{medium-\textbf{u}ncorrelated (deterministic) free path origins }} \\
      \hline
      $s$ & {\small distance since last surface/boundary or uncorrelated birth} & \\
      \hline
      $\Sigma_{tu}(s)$ & {\small uncorrelated macroscopic cross section} & $\Sigma_{tu}(s) = \frac{p_u(s)}{X_u(s)}$ \\
      \hline
      $p_u(s)$ & {\small uncorrelated (equilibrium) free-path distribution} & $p_u(s) = \Sigma_{tu}(s) \e^{-\int_0^s \Sigma_{tu}(s') ds'} = - \frac{\partial}{ \partial s} X_u(s) = \frac{X_c(s)}{\mfp_c}$ \\
      \hline
      $X_u(s)$ & {\small uncorrelated-origin transmittance} & $X_u(s) = 1 - \int_0^s p_u(s') ds'$ \\
      \hline
      $\su$ & {\small mean uncorrelated free-path} & $\su = \int_0^\infty p_u(s) \, s \, ds$ \\
      \hline

    \end{tabular} 
    }   
    \caption{\label{tab:notation}Summary of our notation and relationships between quantities in RGRT.}
  \end{table*}

  Computing the extinction of $p_c(s)$ as normal~\cite{larsen11}, we next find that the attenuation law between collisions is
  \begin{equation}
    X_c(s) = \frac{1}{\sigmean} \left( \frac{r_+ - \sigtilde}{r_+ - r_-} r_+ \e^{-r_+ s} + \frac{\sigtilde - r_-}{r_+ - r_-} r_- \e^{-r_- s} \right).
  \end{equation}
  The non-classical macroscopic cross-section for flights leaving a collision is thus the ratio~\cite{larsen11}
  \begin{equation}
    \Sigma_{tc}(s) = \frac{p_c(s)}{X_c(s)} = \frac{(r_+ - \sigtilde)r_+^2 \e^{-r_+ s} + (\sigtilde - r_-)r_-^2 \e^{-r_- s}}{(r_+ - \sigtilde)r_+ \e^{-r_+ s} + (\sigtilde - r_-)r_- \e^{-r_- s}}
  \end{equation}

  From $\Sigma_{tc}(s)$ we find that the collision probability per incremental path length at the beginning of a path that follows a collision is higher than in the case of no correlation,
  \begin{equation}
    \Sigma_{tc}(0) = \sigmean + \frac{\beta}{\sigmean} \ge \sigmean.
  \end{equation}
  This is intuitively satisfying in the case that one phase is much less dense than the other.  It is more likely to end a collision in the denser phase and so it is more likely to leave a collision from a location in the denser phase, and therefore, the scattering centers are clustered around your current location, producing a short-range collision probability that exceeds the homogeneous prediction.  The atomic mix collision probability density is only encountered at one distance from the collision,
  \begin{equation}
    \Sigma_{tc}\left( \frac{\log \left(\frac{r_+}{r_-}\right)}{\sqrt{4 \beta +(\sigmean-\sigtilde)^2}} \right) = \sigmean
  \end{equation}
  For a long path that avoids collision for large $s$ the cross-section approaches
  \begin{equation}
    \lim_{s \rightarrow \infty} \Sigma_{tc}(s) = r_-.
  \end{equation}

  The uncorrelated-origin free-path distribution, used to enter the medium, is given simply by
  \begin{equation}
    p_u(s) = -\frac{\partial}{\partial s} X_u(s) = \sigmean X_c(s).
  \end{equation}
  The related cross section for uncorrelated origins is the ratio
  \begin{equation}
    \Sigma_{tu}(s) = \frac{p_u(s)}{X_u(s)} = \frac{(r_+ - \sigtilde)r_+ \e^{-r_+ s} + (\sigtilde - r_-)r_- \e^{-r_- s}}{(r_+ - \sigtilde) \e^{-r_+ s} + (\sigtilde - r_-) \e^{-r_- s}}.
  \end{equation}
  The collision probability density beginning an uncorrelated walk uses
  \begin{equation}
    \Sigma_{tu}(0) = \sigmean.
  \end{equation}
  This provides additional support for our conjecture~\cite{deon2018reciprocalii} that the macroscopic cross-section for beginning an uncorrelated walk with all possible realizations in equilibrium has the homogeneous atomic-mix value and is invariant to the spatial correlation imposed on the scattering centers.

  Concurrently, Frankel~\shortcite{frankel2019solution} has derived a $\Sigma_t(s)$ cross-section that corresponds to our $\Sigma_{tu}(s)$, proposing to use this for intercollision statistics, which will lead to nonreciprocal transport and, as we show in Figure~\ref{fig:MC}, does not accurately estimate the free-path lengths between collisions.

  \subsection{Diffusion Approximations}

  We can compute the low-order moments of $p_c(s)$ to determine the various moment-preserving diffusion approximations in the binary mixture in various dimensional spaces (assuming isotropic random media as opposed to layered anisotropic random media of alternating slabs of the two phases).  We easily find
  \begin{align}
    &\ss = \int_0^\infty p_c(s) s^2 ds = \frac{1}{\sigmean} \frac{2 \sigtilde}{\sigtilde-\beta} \\
    &\sss = \int_0^\infty p_c(s) s^3 ds = \frac{1}{\sigmean} \frac{6 \left(\beta +\sigtilde^2\right)}{(\beta -\sigtilde)^2} \\
    &\ssss = \int_0^\infty p_c(s) s^4 ds = \frac{1}{\sigmean} \frac{24 \left(\beta  (2 \sigtilde+1)+\sigtilde^3\right)}{(\sigtilde-\beta )^3}
  \end{align}
  from which the Green's function approximations follow directly~\cite{deon2018reciprocalii}.  We briefly note that the diffusion coefficient for collision rate density about an isotropic point source that emits from the scattering centers in the medium is
  \begin{equation}
    D_{C_c} = \frac{\ss}{2 d} = \frac{1}{\sigmean} \frac{ \sigtilde}{d(\sigtilde-\beta)}
  \end{equation}
  where $d$ is the dimension of the space in which scattering occurs.  This reduces to the classical $P_1$ result of $D = 1/(d \sigmean)$ when the cross sections in both phase match and $\beta = 0$.  The effective attenuation coefficient for scalar collision rate density about a correlated-emission point source is
  \begin{equation}
    \Sigma_{\text{eff}} = \sqrt{ \frac{d (1-c)  \sigmean ( \sigmean \sigtilde-\beta)}{\sigtilde} }
  \end{equation}
  and for uncorrelated emission or scalar flux about a correlated emitter is
  \begin{equation}
    \Sigma_{\text{eff}} =  \sqrt{ \frac{(c-1) d \sigmean (\beta -\sigmean \sigtilde)^2}{\beta 
   (c-1) \sigmean+\beta  c \sigtilde-\sigmean \sigtilde^2} }.
  \end{equation}

  \subsection{Half-Space Single-Scattering Reflection Law}
  
    We now consider a plane-parallel problem in a 3D half-space with isotropic scattering and index-matched smooth boundary.  The BRDF $f_1$ for single scattering is given by~\cite{deon2018reciprocali}
    \begin{multline}
      f_1(\mu_i,\mu_o) = \frac{1}{\mu_i \mu_o} \frac{c}{4 \pi} \frac{1}{\mfp_c} \int_0^\infty  X_c \left(\frac{z}{\mu_i}\right) X_c \left(\frac{z}{\mu_o}\right) dz = \\
      \frac{c }{4 \pi (\mu_i+\mu_o)} \left(1-\frac{\beta  (\sigtilde+1) \mu_i \mu_o}{(\mu_i^2+\mu_o^2)(\sigtilde-\beta )+\mu_i \mu_o \left(2 \beta +\sigtilde^2+1\right) }\right)
    \end{multline}
    where $\mu_i$ is the cosine of the inclination to the normal vector of the half space for the incident illumination, and $\mu_o$ is the outgoing cosine.  We see the classical result scaled by a factor that decreases the reflectance, and when the cross sections of both phase match and $\beta \rightarrow 0$, the classical result is recovered, as desired.  This results appears to be new.
    Note, again, that this derivation assumes an isotropic multidimensional random mixture~\cite{larmier2016finite,larmier2017benchmark}, not a 3D volume with layered slabs separated by exponential chord lengths in the direction of the normal, for which results such as these are known~\cite{pomraning1988classic}.  

  \section{Binomial Binary Mixtures}

    Three distinct models for transport in one-dimensional stochastic random media with densities described by a binary binomial random process have been proposed by Williams~\shortcite{williams1997radiation} and Akcasu and Williams~\shortcite{akcasu2004analytical}.  We set out to derive the form of RGRT that corresponds to each of the three models but, much to our surprise, found that the three attenuation laws are, in fact, exponential laws in disguise. We briefly present the homogenization relations for the three models, which are exact.  These results are easily verified in software packages such as Mathematica by testing that the attenuation laws are proportional to their derivatives.

    \subsection{Model 1}

      This model considers stacks of slabs, each of uniform thickness $d > 0$, where the total cross sections are chosen independently and randomly from $\Sigma_0 - \sigma$ and $\Sigma_0 + \sigma$, with equal probability (and $\sigma < \Sigma_0$).  The stationary interpretation of the attenuation law $X_u(s)$ is~\cite{williams1997radiation}
      \begin{equation}\label{eq:X1}
      	X_u(s) = e^{-s \Sigma_0 } \cosh ^{(s/d)}(d \sigma ).
      \end{equation}
      Equation~\ref{eq:X1} can be expressed as an exponential of $s$,
      \begin{equation}
      	X_u(s) = \e^{ -\bar{\Sigma} s }
      \end{equation}
      with effective attenuation coefficient
      \begin{equation}
      	\bar{\Sigma} = \Sigma_0 + \frac{\log (\text{sech}(d \sigma ))}{d}.
      \end{equation}

    \subsection{Model 2}

      This model considers two different materials in alternating slabs.  The cross sections are $\Sigma_1 = \Sigma_{01} + \beta \sigma_1$ and $\Sigma_2 = \Sigma_{02} + \alpha \sigma_2$, where $\alpha$ and $\beta$ are random variables which take on values $\pm 1$.  The slab thicknesses are also allowed to differ, $d_1$ and $d_2$, respectively.  The stationary interpretation of the attenuation law $X_u(s)$ is~\cite{williams1997radiation}
      \begin{equation}\label{eq:X2}
        X_u(s) = 2^{-s} \left(e^{d_1 \sigma_1+d_2 \sigma_2}+1\right)^s
   e^{\frac{1}{2} s (-d_1 (\Sigma_{01}+\sigma_1)-d_2
   (\Sigma_{02}+\sigma_2))}.
      \end{equation}
      Equation~\ref{eq:X2} can be expressed as an exponential of $s$,
      \begin{equation}
        X_u(s) = \e^{ -\bar{\Sigma} s }
      \end{equation}
      with effective attenuation coefficient
      \begin{equation}
        \bar{\Sigma} = \frac{1}{2} \left(-2 \log \left(e^{d_1 \sigma_1+d_2 \sigma_2}+1\right)+d_1 (\Sigma_{01}+\sigma_1)+d_2
   (\Sigma_{02}+\sigma_2)+\log (4)\right).
      \end{equation}

    \subsection{Model 3}

      Model 3 includes variable thickness of the slabs, $\delta_i = \bar{\delta} + \alpha_i \Delta$ being the thickness of each slab, with $\alpha_i$ random variables taking on $\pm 1$, and the same cross-section in each slab $\Sigma_0$.  The proposed attenuation law is~\cite{williams1997radiation}
      \begin{equation}\label{eq:X3}
        X_u(s) = e^{-s \Sigma_0} \cosh ^{\frac{s}{d}}(\sigma  (\Delta +\bar{\delta}))
      \end{equation}
      Equation~\ref{eq:X3} can be expressed as an exponential of $s$,
      \begin{equation}
        X_u(s) = \e^{ -\bar{\Sigma} s }
      \end{equation}
      with effective attenuation coefficient
      \begin{equation}
        \bar{\Sigma} = \Sigma_0-\frac{\log (\cosh (\sigma  (\Delta +\bar{\delta})))}{d}.
      \end{equation}

\section{Conclusion}

  We have taken the Levermore-Pomraning attenuation law for Markovian stochastic binary mixtures and derived the related form of RGRT that encompasses this law.  The derivation has exhibited the free-path distribution between collisions, a new result that appears to be exact for forward scattering in a rod.  The complete RGRT formulation deterministically unifies the general theories of GRT~\cite{rybicki1965transfer,peltoniemi1993radiative,davis06,moon07,taine2010generalized,davis14,davis18} and nonclassical Boltzmann transport~\cite{larsen11} with the Levermore-Pomraning theory~\cite{pomraning98} and the related nonexponential random flight Monte Carlo acceleration scheme of Audic and Frisch~\shortcite{audic1993monte}, while avoiding the need for any accumulation of free-path statistics in a preprocess.

  In our derivation we took the Levermore-Pomraning law as a black-box input with abstract parameters.  No knowledge of the derivation of the law or the meaning of the parameters took any role in the subsequent analysis, yet known properties about the microstructure naturally fell out of the process, such as the atomic mix macroscopic cross section, which appears at the beginning of a free-path with a deterministic origin,
  \begin{equation}
    \Sigma_{tu}(0) = \sigmean,
  \end{equation}
  providing additional motivation for a conjecture in the previous paper in this series~\cite{deon2018reciprocalii}.  This exercise illustrates how the transport distributions in GRT, which can be arrived at in ways that are quite disconnected from knowledge of the microstructure (or may correspond to laws for which no microstructure could ever exhibit), contain within them additional information about the microstructure.

  One possible interpretation of our derivations is that we have transformed the binary mixture of the two phases, each with particles in them, into a two phase medium with one phase void and the other corresponding to the scatterers alone.  The double heterogeneity is collapsed, and the transmission law relates to the lineal path function~\cite{torquato93} in the void phase, from which we immediately produce the chord-length distribution, which becomes the inter-collision free-path distribution in the limit that the particles shrink to zero size and infinite density to produce classical exponential attenuation in each phase.  The power of this relationship should also prove useful in more complex mixtures and layers of heterogeneity where, by reciprocity, we expect this relationship to hold.

  Important future work includes expanding upon the benchmarks of Audic and Frisch~\shortcite{audic1993monte} to compare the accuracy of RGRT for binary mixtures, using recent benchmark solutions for multidimensional mixtures~\cite{larmier2017monte,larmier2018monte}, and also to measure efficiency relative to the CLS algorithm.

  While GRT holds promise for binary mixtures, the current presentation is limited to nonstochastic single-scattering albedo and phase function, assumptions long known to greatly simplify the solutions.  It is not possible to support different absorption levels in the two phases.  Nor can stochastic reaction rates be estimated (such as the collision rate in phase 2, for .e.g, or, by reciprocity, a volume source in the medium that emits from the scatterers in phase 2).  These restrictions were not acknowledged by Audic and Frisch and are important to clarify.  Future work is required to generalize RGRT to path-length dependent single-scattering albedo $c(s)$ and derive this function for Markovian binary mixtures.

  It is also worth mentioning that, given $p_c(s)$ for Markovian binary mixtures, we can relate transport in such media to nonexponential random flights to apply the Cauchy-like formulas for mean total track-length and collision rate inside of finite volumes under uniform illumination, that have been generalized to provide results at arbitrary positions in phase space~\cite{mulatier2014universal}.

  In the next paper, where plane-parallel transport in RGRT is explored under a variety of free-path distributions, we will expand upon the single-scattering BRDF in this paper to produce complete solutions for half space transport with isotropic scattering.

\bibliographystyle{acmsiggraph}
\bibliography{nonexppartiii}

\end{document}